\documentclass[prb, showpacs, twocolumn, aps, superscriptaddress, a4paper]{revtex4-1}
\usepackage{dcolumn, amssymb, amsmath, amsfonts, graphicx, latexsym, color, braket, subfigure}
\usepackage{epstopdf}
\DeclareMathOperator{\Tr}{Tr}
\DeclareMathOperator{\sech}{sech}
\begin{document}

\title{Critical behavior of order parameter at the nonequilibrium phase transition of the Ising model}
\author{Bin Li}
\affiliation{Department of Physics, Zhejiang Normal University, Jinhua 321004, People's Republic of China}
\author{Chao Gao}
\affiliation{Department of Physics, Zhejiang Normal University, Jinhua 321004, People's Republic of China}
\author{Gao Xianlong}
\affiliation{Department of Physics, Zhejiang Normal University, Jinhua 321004, People's Republic of China}
\author{Pei Wang}
\email{wangpei@zjnu. cn}
\affiliation{Department of Physics, Zhejiang Normal University, Jinhua 321004, People's Republic of China}
\date{\today}

\begin{abstract}
After a quench of transverse field, the asymptotic long-time state of Ising model
displays a transition from a ferromagnetic phase to a paramagnetic phase as the post-quench field strength increases,
which is revealed by the vanishing of the order parameter defined as the averaged magnetization over time.
We estimate the critical behavior of the magnetization at this nonequilibrium phase transition by using mean-field approximation.
In the vicinity of the critical field, the magnetization vanishes as the inverse of
a logarithmic function, which is significantly distinguished from the critical behavior of
order parameter at the corresponding equilibrium phase transition, i.e. a power-law function.
\end{abstract}

\maketitle

\section{\label{sec:level1}Introduction}
\begin{figure*}
\includegraphics{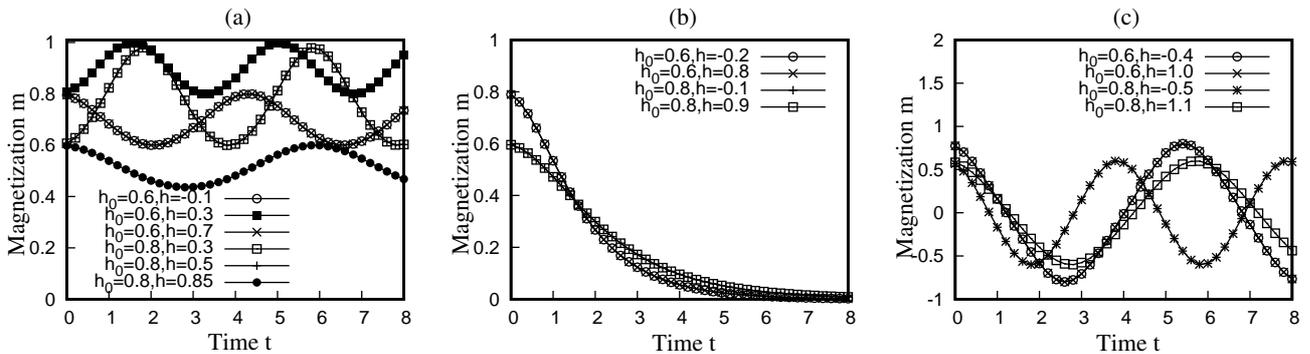}
\caption{\label{fig:wide} The evolution of the magnetization for different values
of $h_0$ and $h$: (a) $h $ is between $h_c^-$ and $h_c^+$, (b) $h=h_c^\pm$,
and (c) otherwise. $m(t)$ is a periodic function in the panels~(a) and~(c),
but decays to zero exponentially in the panel~(b). The magnetization never
changes sign in the panel~(a), but oscillates around zero in the panel~(c).
Some curves overlap with each other, revealing a symmetry in
the dynamics.}
\end{figure*}

The properties of a quantum many-body system out of equilibrium are
attracting more and more attention in recent years~\cite{Polkovnikov11,Eisert15}.
Similar to the equilibrium states, the properties of a nonequilibrium state
can also exhibit some abrupt change, which leads to different notions of nonequilibrium phase transition.
Among them, there is the steady-state phase transition~(SPT), defined by the nonanalyticity
of some observable as a function of the parameters of the nonequilibrium protocol in the
asymptotic long-time states~\cite{Diehl08,Diehl10,Sciolla10,Wang16b}.
The dynamical phase transition~(DPT) is signaled by
an abrupt change of the way of a physical quantity relaxing to its asymptotic value~\cite{Eckstein09},
which can be an exponential relaxation, a power-law relaxation, or no relaxation (everlasting
oscillations)~\cite{Foster14}. Finally, the dynamical quantum phase transition (DQPT) is given by the
nonanalyticity of the dynamical free energy as a function of time in the transient
and the intermediate time scale~\cite{Heyl13}.
The connection between these different notions was addressed recently~\cite{Halimeh17,Zunkovic18,Wang18}.

Similar to the phase transitions in equilibrium states, a nonequilibrium SPT can be described by
a local order parameter or a topological one, depending on whether the transition
breaks some symmetry or is topologically driven.
However, in nonequilibrium states,
no function plays the role of free energy whose nonanalyticity determines
the nonanalyticity of all the other quantities in equilibrium states.
The nonanalyticity at a SPT
 is assigned to certain observable,
 and can be much different from its equilibrium counterpart.
An example is the Hall conductance at a topologically driven SPT, which changes continuously
with a logarithmically-divergent derivative~\cite{Wang16b}.
While at the corresponding ground-state phase transition,
 the Hall conductance jumps from one plateau to another.

The transverse Ising model~(TIM) is a prototypical example for studying the symmetry-breaking phase transitions.
In the case of one dimension and short-range interaction, the model can be solved strictly by
the Jordan-Wigner transformation~\cite{Lieb61,Pfeuty70,Barouch71}.
It undergoes a transition from a ferromagnetic phase in weak field to a paramagnetic phase in strong field~\cite{Sachdev99}.
The quench dynamics of TIM was estimated~\cite{Sengupta04}, both for the local observables
and the entanglement~\cite{Amico08,Dutta15}.
Beyond one dimension, TIM has no exact solution,
 and the mean-field approximation was usually taken for studying the phase transition in equilibrium states,
 even if it fails to predict the correct critical exponent in two and three dimensions.
 On the other hand, TIM with infinite-range interaction,
 which is equivalent to the Lipkin-Meshkov-Glick (LMG) model~\cite{Lipkin65a,Lipkin65b,Lipkin65c},
is also exactly solvable with the solution fitting well into the mean-field picture.
The LMG model undergoes a second order phase transition at a critical field~\cite{Botet82}.
It caused revived interest in studying the relation between quantum phase transition
and entanglement entropy~\cite{Vidal07,Amico08}.
The quench dynamics of the LMG model was exploited by Das et al.~\cite{Das2006},
 and the generalization to other models with infinite-range interaction was made by Sciolla and Biroli~\cite{Sciolla10,Sciolla11}.
 There exists a similar ferromagnetic-paramagnetic phase transition in the asymptotic state of the LMG model after a quench~\cite{Sciolla11},
corresponding to the transition in the ground state.
Such a phase transition was also found in the one-dimensional TIM with power-law interactions~\cite{Halimeh16}.

The nonequilibrium dynamics of TIM both in one dimension and with infinite-range interaction
has been intensively studied, but the critical behavior of the magnetization
at the corresponding ferromagnetic-paramagnetic transition has not been clearly addressed.
In this paper, we estimate the critical behavior of the magnetization in the asymptotic long-time state of TIM
by using the mean-field approximation.
We find that the magnetization vanishes as the inverse of a logarithmic function at the critical field,
which is qualitatively different from the critical behavior at the corresponding equilibrium phase transition.
The latter is well known to follow a power-law form.
It is worth mentioning that this logarithmic singularity has already been found in the Fermi-Hubbard model~\cite{schiro10},
the Bose-Hubbard model~\cite{Sciolla10} and the $\phi^4$ $N$-components field theory~\cite{Sciolla13b},
and supposed to be a general feature of the mean-field quench dynamics.
But it has not been directly observed yet. This paper provides a proof of the logarithmic
singularity in the LMG model, which can be realized
in trapped ions~\cite{suzuki2012quantum, gordon1999creating}, and then contributes
to the possible observation of this logarithmic singularity in future. 

To keep our paper self-consistent, we review the mean-field theory of TIM
for both the ground state and the quench dynamics in Sec.~\ref{sec:GS}.
In Sec.~\ref{sec:EXP}, we show that the quench dynamics of TIM in mean-field approximation
is equivalent to that of the LMG model.
In Sec.~\ref{sec:CBOP}, we prove the logarithmic critical behavior by both analytical and
numerical methods and compare it with
the equilibrium counterpart. Section~\ref{sec:CONC} summarizes our results.

\section{\label{sec:GS}Mean-field quench dynamics}
The Hamiltonian of TIM is
\begin{equation}\label{eq:isingm}
\hat{H}=-\frac{J}{2d}\sum_{\langle i, j\rangle}\hat{\sigma}_{i}^{z}\hat{\sigma}_{j}^{z}+h\sum_{i}\hat{\sigma}_{i}^{x},
\end{equation}
where $J>0$ is the ferromagnetic coupling between spins.
$d$ denotes the spatial dimension of the model
with $2d$ representing the adjacent number of each lattice site.
It appears in the denominator of the coupling strength in order to normalize the energy density.
$\hat{\sigma}_{x}$ and $\hat{\sigma}_{z}$ are the Pauli matrices and $h$
is the magnetic field along the direction $x$.
$\langle i, j\rangle$ is the index of two lattice sites of nearest neighbors.
We first briefly review the case when the system is in equilibrium with the transverse field being a constant $h=h_0$.
In the mean field theory, we replace the Pauli matrix $\hat{\sigma}^z_j$ with
its expectation value $\langle \hat{\sigma}_j^z\rangle$=$m_0$.
For convenience, we take $J$ as the energy unit, thereafter, the effective mean-field
Hamiltonian is expressed as $\hat{H}_{\text{eff}}=-m_0 \sum_{i} \hat{\sigma}_{i}^{z}+h\sum_{i}\hat{\sigma}_{i}^{x}$.
In equilibrium states, the magnetization of the system with $\left| h_0 \right| <1$ is expressed as
\begin{eqnarray}
\begin{aligned}
m_0&=\frac{\Tr[e^{-\beta\hat H_{\text{eff}}}\hat{\sigma}_{i}^{z}]}{\Tr[e^{-\beta{\hat{H}_{\text{eff}}}}]}\\
&=\frac{m_0}{\sqrt{m_0^2+h_0^2}}\tanh \left(\beta{\sqrt{m_0^2+h_0^2}}\right),
\end{aligned}
\end{eqnarray}
where $\beta=1/(k_{B}T)$ and $k_{B}$ is the Boltzmann constant.
$T$ is the temperature of the system, which tends to $0$ when the system is in the ground state.
The magnetization and the magnetic field of the ground state satisfy
\begin{equation}\label{eq:m0h0}
\sqrt{m_0^2+h_0^2}=1.
\end{equation}
We can also solve the expectation value of $\hat \sigma^x_j$ and $\hat \sigma^y_j$
in the ground state, which are $m_0^x=-h_0$ and $m_0^y=0$, respectively.
For the ground state with $\left| h_0\right| > 1$,
the expectation values of $\hat \sigma^x_j$, $\hat \sigma^y_j$
and $\hat \sigma^z_j$ are $m_0^x=-1$, $m_0^y=0$ and $m_0=0$, respectively.

\begin{figure}[b]
  \includegraphics{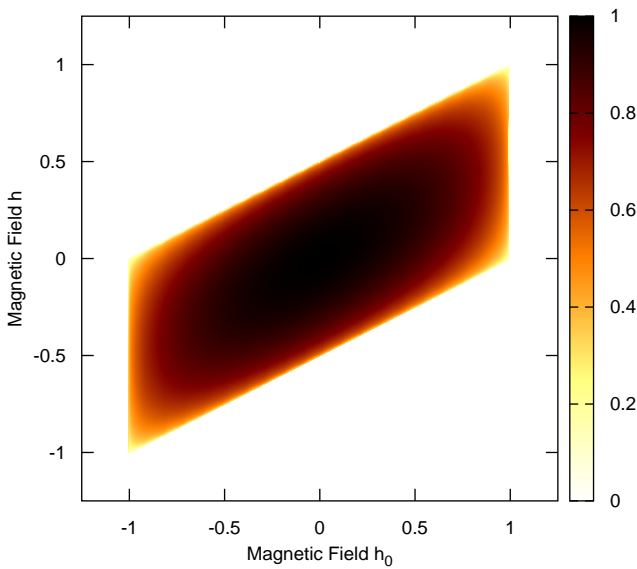}
  \caption{\label{fig:Mh0h} The averaged magnetization $M$ in the long time limit
  in the $h_0-h$ plane. The colored region is the ferromagnetic phase with nonzero $M$,
  but the white region is the paramagnetic phase with $M=0$.}
\end{figure}

Suppose that the system is initially prepared in the ground state of $h_0$, and then
quenched at the time $t=0$ by suddenly changing the transverse field from $h_0$ to $h$.
In the spirit of mean-field approximation, the dynamics of the wave function is governed
by an time-dependent effective Hamiltonian $\hat{H}_{\text{eff}}(t)=-m(t)\sum_{i}
\hat{\sigma}_{i}^{z}+h\sum_{i}\hat{\sigma}_{i}^{x}$, where $m(t)$ is the expectation value
of $\hat \sigma^z_j$ at the time $t$.
The time-dependent expectation values of $\hat \sigma^x_j$, $\hat \sigma^y_j$
and $\hat \sigma^z_j$ satisfy a system of differential equations:
\begin{eqnarray}\label{eq:partial}
\begin{aligned}
\frac{dm_x}{dt}&=2mm_y ,\\
\frac{dm_y}{dt}&=-2mm_x-2hm, \\
\frac{dm}{dt}&=2hm_y.
\end{aligned}
\end{eqnarray}
Since the initial state at $t=0$ is the ground state at $h_0$, the initial conditions for solving
Eq.~(\ref{eq:partial}) are $m_x(0)=m_0^x$, $m_y(0)=0$, $m(0)=m_0$.
It is easy to eliminate $m_x$ and $m_y$ in Eq.~(\ref{eq:partial}) and obtain for
$\left| h_0 \right| <1$
\begin{eqnarray}\label{eq:magtime}
\begin{aligned}
  \left(\frac{dm}{dt}\right)^2=-(m^2-m_0^2)\left[m^2+4(h-h_c^-)(h-h_c^+)\right],
\end{aligned}
\end{eqnarray}
where $h_c^{\pm}=(h_0\pm 1)/2$.

The dynamics of magnetization depends on both the initial and the post-quench magnetic fields.
As
$\left| h_0 \right|>1$,
the magnetization is always zero.
As $\left| h_0 \right|<1$, we find three different patterns in the dynamics of magnetization,
as shown in Fig.~\ref{fig:wide}.
At the critical point $h=h_c^\pm$, [see Fig.~\ref{fig:wide}(b)],
the solution of Eq.~(\ref{eq:magtime}) is found to be $m(t)=m_0\sech(m_0 t)$,
which decays to zero exponentially.
As $h$ is between $h_c^-$ and $h_c^+$, the magnetization has an everlasting
oscillation, but its sign never changes~[see Fig.~\ref{fig:wide}(a)].
The value of $m(t)$ is always beyond $m_0$ for $0<h<h_0$,
but below $m_0$ for $h_c^-<h<0$ or $h_0<h<h_c^+$.
Furthermore, Eq.~(\ref{eq:magtime}) has the exactly same solution
for two different values of $h$ which are symmetric with respect to $h_0/2$.
This symmetry in the dynamics of magnetization is shown in Fig.~\ref{fig:wide}(a)
(see the matching curves). Finally, as $h<h_c^-$ or $h>h_c^+$,
$m(t)$ also displays an everlasting oscillation,
 but this oscillation is exactly symmetric about $m=0$
 [see Fig.~\ref{fig:wide}(c)].
The average of magnetization over one period is zero.

Since $m(t)$ is a periodic function of time for $h \neq h_c^\pm$,
we can use the averaged magnetization
 $M=\int_0^{\tau} m(t) dt/\tau$
  over one period $\tau$ as the order parameter.
The region $h_c^- <h <h_c^+$ is then in the ferromagnetic phase
since the averaged magnetization is nonzero,
 while the regions $h<h_c^-$ and $h>h_c^+$
are in the paramagnetic phase since the averaged magnetization is zero.
The two points $h_c^-$ and $h_c^+$ are the critical points of
the ferromagnetic-paramagnetic phase transition.
This phase transition must be distinguished from an equilibrium phase transition,
since the magnetization does not thermalize in the mean field theory.
Alternatively, this phase transition should be viewed as a SPT.
SPT usually means the nonanalyticity in the steady limit which an observable relaxes to.
In the mean field theory, the magnetization does not relax,
but keeps on oscillating.
But we can re-express the averaged magnetization in one period as
\begin{equation}
M = \lim_{T\to \infty} \frac{1}{T} \int^T_0 dt  \, m(t).
\end{equation}
This average of an observable in the long time limit must be equal
to the steady limit of an observable if it exists. Therefore,
we can generalize the definition of SPT to the nonanalyticity of
the averaged observable in the long time limit, in which sense
the ferromagnetic-paramagnetic phase transition discussed in this paper
is a SPT.

We find no simple explicit expression of $m(t)$ by solving Eq.~(\ref{eq:magtime}).
While we can still obtain a tidy expression for the period of $m(t)$,
which serves as the basis for analyzing the critical behavior of $M$.
The period when $h_c^-<h<h_c^+$ can be expressed as
\begin{equation}
\tau = \frac{4K}{B},
\end{equation}
where
\begin{eqnarray}\label{eq:BKk}
\begin{aligned}
B &=\sqrt{2m_{0}^{2}+4hh_{0}-4h^{2}+2m_{0}\sqrt{m_{0}^{2}+4(hh_{0}-h^{2})}},\\
K &=\frac{\pi}{2}\left\{\sum\limits_{n}\left[\frac{(2n-1)!!}{2^{n}n!}\right]^{2}k^{2n}+1\right\}, \\
k &=\sqrt{\frac{2m_{0}^{2}+4hh_{0}-4h^{2}-2m_{0}\sqrt{m_{0}^{2}+
4(hh_{0}-h^{2})}}{2m_{0}^{2}+4hh_{0}-4h^{2}+2m_{0}\sqrt{m_{0}^{2}+4(hh_{0}-h^{2})}}}.
\end{aligned}
\end{eqnarray}
To obtain Eq.~(\ref{eq:BKk}), we have used the fact that $K$ is the first type of elliptic integral.
As a function of $k$, $K$ diverges as $ k \to 1 $.
At the same time,
$k$ is a function of $h$, and its value at the critical
magnetic fields is $ k (h_c^\pm) = 1 $. Therefore,
the period $\tau$ diverges as $h$ approaches $ h_c^\pm $.

The averaged magnetization is found to be
\begin{eqnarray}\label{eq:oph}
M=\frac{\pi B}{4K},
\end{eqnarray}
which is the inverse of $ \tau $, thereafter, it vanishes
as $h$ approaches $ h_c^\pm $.
Figure~\ref{fig:Mh0h} exhibits
$M$ in the $h_0 - h$ plane, in which we can easily read out
the different phases, i.e., the colored region with nonzero $M$ is the ferromagnetic phase
while the purely white region is the paramagnetic phase.

\begin{figure}[hbt]
  \includegraphics{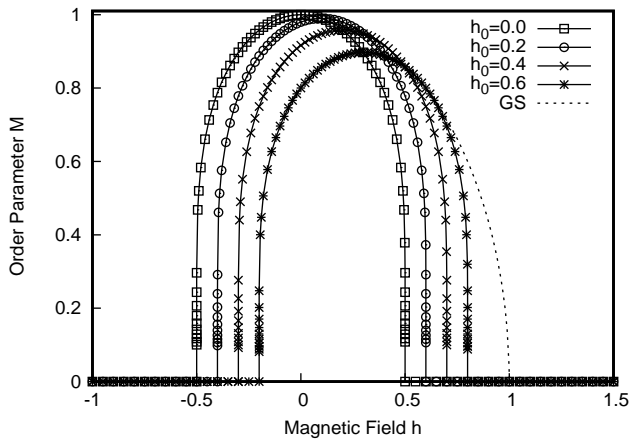}
  \caption{\label{fig:Mh1} The solid lines with points represent the averaged magnetization
  after a quench as a function of $h$ for different values of initial magnetic field $h_0$.
  The curve of  $M(h)$ is always symmetric with respect to $h=h_0/2$.
While the dashed line, following the expression $\sqrt{h^2+M^2}=1$,
 represents the magnetization of the ground-state with constant magnetic field $h$.}
\end{figure}

Figure~\ref{fig:Mh1} displays how $M$
varies as a function of $h$
  for different values of $h_0$.
The maximum of the magnetization
$M_{\text{max}}=(\pi/4)(1+m_0)/K[(1-m_0)/(1+m_0)]$
 decreases with $h_0$ increasing.
It would be interesting to compare $M(h)$ with the ground-state magnetization
as a function the magnetic field (see the dashed line of Fig.~\ref{fig:Mh1}).
As we expect, the magnetization in the ground state significantly deviates from that in the nonequilibrium case, 
as $h$ goes away from $h_0$,
 because the system is far from thermal equilibrium after a quench.
And $M(h)$ drops to zero more sharply than the dashed line,
 indicating the presence of an exotic critical behavior, which will be discussed in Sec.~\ref{sec:CBOP}.

\section{\label{sec:EXP}Comparison between TIM in the mean field approximation
and the LMG model}

In this section we consider the applicability of the mean field approximation.
In the one-dimensional TIM, the mean field approximation does not work due to the strong quantum fluctuation,
 and it is useless since the one-dimensional TIM is strictly solvable by the Jordan-Wigner transformation.
For dimensions higher than one, TIM is nonintegrable, and thus thermalizes in the long-time limit,
 contradicting the prediction of the mean field theory.
Therefore, the mean-field theory fails to describe how the steady limit of magnetization changes with the quenched magnetic field $h$,
 alternatively, which should be determined by the Gibbs ensemble with an $h$-dependent generalized temperature.
But it is worth mentioning that the mean-field approximation is usually believed
to work in the intermediate time scale for high dimensions.
If this is true in our system,
 Eq.~(\ref{eq:oph}) gives the magnetization in the quasi-stationary states of TIM after quench,
  which will finally change into a thermalized state in the long-time limit.
Anyway, there does exist another model in which the magnetization exactly follows Eq.~(\ref{eq:oph})
 and that can be realized in trapped ions~\cite{suzuki2012quantum, gordon1999creating}.
 It is the LMG model.

Next we shortly introduce the dynamics of
the LMG model by following Ref.~[\onlinecite{Sciolla11}]. The Hamiltonian of the LMG model is written as
\begin{equation}\label{eq:LMGm}
\hat H_{\text{LMG}}= - \frac{J}{N} \sum_{i>j} \hat \sigma_i^z \hat \sigma_j^z + h \sum_j \hat \sigma_j^x,
\end{equation}
where $N$ denotes the total number of spins.
The interaction between different spins is a constant, which does not decay as the distance between spins increases.
 A fully-connected model like Eq.~(\ref{eq:LMGm}) is invariant under permutation of spins.
 This symmetry reduce the many-body Hilbert space into subspaces of dimension $N$,
  in which the dynamics of the magnetization is described by a single-particle Schr\"{o}dinger equation.
  In the thermodynamic limit $N\to \infty$,
this Schr\"{o}dinger equation can be replaced by a classical Hamilton's equation which is
\begin{eqnarray}\label{eq:mp}
\begin{aligned}
  \dot{m}&= - 2h\sqrt{1-m^{2}} \sin p\\
  \dot{p}&=2m+h\frac{2m}{\sqrt{1-m^{2}}}\cos p,
\end{aligned}
\end{eqnarray}
where $m$ is the magnetization and $p$ is the canonical momentum corresponding to $m$.
Notice that the energy of the system $E=-m^2+2h\sqrt{1-m^2}\cos p$
is conserved during an evolution, we can then replace $p$ by a function of $m_0$ and $m$.
The result is exactly Eq.~(\ref{eq:magtime}).

In summary, the dynamics of the magnetization is the same for the model~(\ref{eq:LMGm})
and the model~(\ref{eq:isingm}) in the mean field approximation.
Therefore, when we discuss the critical behavior of the magnetization following Eq.~(\ref{eq:oph}),
our results can indeed be tested by an experiment simulating the LMG model.

\section{\label{sec:CBOP}The critical behavior of the order parameter}

The averaged magnetization~(\ref{eq:oph}) serves as the order parameter for
the ferromagnetic-paramagnetic phase transition in the quenched state of the TIM or the LMG model.
The critical behavior of the order parameter
is usually a focus of attention in the study of continuous phase transition in thermal equilibrium,
since it indicates the universality class of the phase transition. Similarly,
we will check the critical behavior of $M$ at this out-of-equilibrium SPT.

In order to extract the behavior of $M(h)$,
as $ h \to h_c^\pm $,
we expand Eq.~(\ref{eq:oph}) in the vicinity of $ h = h_c^\pm $ (the ferromagnetic side)
 as a function of $\Delta h=\left| h - h_c^\pm \right|$,
which denotes the distance to the SPT point.
First, the denominator $B(h)$ and the quantity $k$ are expanded saprately to be
$ B (h) =m_0+ 2 \Delta h^{1/2} + \mathcal{O} [(\Delta h) ^ {3/2}]$ and $ k = 1-4/m_0 \Delta h^{1/2}+ \mathcal{O} (\Delta h) $,
where $\mathcal{O}$ is the big-O notation.
While the quantity $K$ in the numerator, can be initially expanded as a function of $1-k$:
\begin{eqnarray}
\begin{aligned}
K(k)=-\frac12 \left[\ln(1-k)-4\ln2\right]+\mathcal{O}(1-k),
\end{aligned}
\end{eqnarray}
 since $k\to 1$ if and only if $ h \to h_c^\pm $.
Obviously, $K$ diverges in a logarithmic way in the limit $k\to 1$,
 and the infinitesimal term $ \mathcal{O}(1-k) $ can be neglected
  compared to the logarithmic divergence as $k$ is close enough to $1$.
Finally, the order parameter $M$ can be expressed as
\begin{eqnarray}\label{eq:critical}
\begin{aligned}
M =-\frac{\pi m_0}{\ln(\Delta h)-2\ln(4m_0)}+\mathcal{O}(\Delta h).
\end{aligned}
\end{eqnarray}
It is easy to see that the first term of Eq.~(\ref{eq:critical}) goes to zero much slower,
 and overwhelms the second term in the limit $\Delta h \to 0$.
Therefore, the critical behavior of $M$ in a compact form is
\begin{equation}\label{eq:logs}
M\sim -\frac{\pi m_0}{\ln(\Delta h)}.
\end{equation}

\begin{figure}[hbt]
  \includegraphics{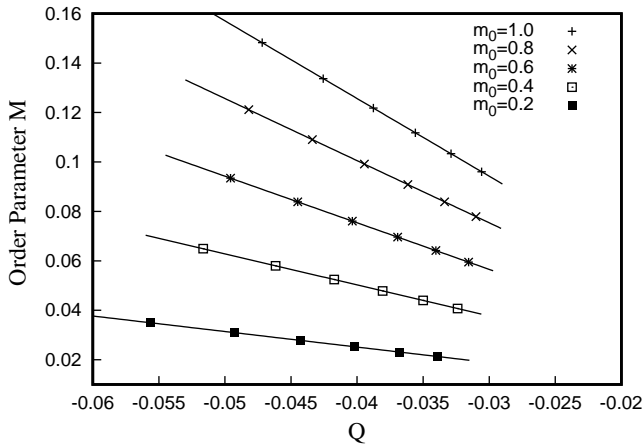}
  \caption{\label{fig:Mh2} The order parameter $M$ as a function of
  $Q=1/[\ln(\Delta h)-2\ln(4m_0)]$ for different values of
  $m_0$ ranging from $0. 2$ to $1. 0$.
  Different kinds of points show the numerical results of $M$, while the solid straight lines denote $M=-m_0\pi Q$.}
\end{figure}

In Fig.~\ref{fig:Mh2}, we show the numerical results of the order parameter $M$
as a function of $Q=1/[\ln(\Delta h)-2\ln(4m_0)]$ in the vicinity of the phase transition for different $m_0$.
The numerics of $M(Q)$ fits well with straight lines of the slope $-m_0\pi$, verifying our analytical result~(\ref{eq:critical}).

As shown in Eq.~(\ref{eq:oph}) and also in Figs.~\ref{fig:Mh0h} and~\ref{fig:Mh1},
the order parameter has a complicated expression
 when the quenched state is far away from the critical point.
While in the vicinity of the critical point, the behavior of the order parameter
simplifies into the inverse of a logarithmic function.
This logarithm-type critical behavior of the order parameter is nontrivial,
 which, up to the best of our knowledge, does not exist in any equilibrium phase transition.
In equilibrium phase transitions,
the order parameter is a power-law function with universal exponent in the vicinity of critical point,
 which reflects the scaling invariance of the system when the correlation length is divergent.
For example,
the ground-state magnetization of TIM in the mean-field theory exhibits
 $m_0 \sim \Delta h^\alpha $ with $\alpha={1/2}$, according to Eq.~(\ref{eq:m0h0}).
Beyond the mean-field approximation, the value of $\alpha$ changes,
but the power-law form $\Delta h^\alpha $ keeps, which is significantly distinguished from Eq.~(\ref{eq:logs}).

Equation~(\ref{eq:logs}) is reminiscent of the critical behavior of the Hall conductance
at a topologically-driven SPT~\cite{Wang16b}.
Both the magnetization and the Hall conductance
are the observables displaying nonanalyticity at the SPTs.
The magnetization vanishes as the inverse of a logarithmic function,
 while the derivative of the Hall conductance diverges as a logarithmic function.
  Both are qualitatively different from the critical behavior at the corresponding equilibrium phase transitions,
   indicating that an exotic critical behavior is always a feature of nonequilibrium SPTs,
   whether the SPT is a topological phase transition or a symmetry-breaking one.

\section{\label{sec:CONC}Conclusion}

In conclusion,
we have studied the ferromagnetic-paramagnetic phase transition
 in the asymptotic long-time state of the TIM and the LMG model after a quench.
Not only the transition point but also the critical behavior of the order parameter are different from their equilibrium counterparts.
The order parameter, defined as the averaged magnetization in the long time limit,
vanishes as the inverse of a logarithmic function at this nonequilibrium SPT,
 which is qualitatively different from the power-law critical behavior in thermal equilibrium.

The dynamics of the magnetization are the same
 in the TIM under mean-field approximation
 and in the LMG model.
An ensemble of trapped ions provides a possible experimental platform
 for testing the exotic critical behavior of magnetization.

\section*{acknowledgments}
This work is supported by NSF of China under Grant Nos.~11604300,~11774315,~11774316
and~11835011.
Pei Wang is also supported by the Junior Associates program
of the Abdus Salam International Center for Theoretical Physics.

During the preparation of the paper,
we notice that parts of the results were discussed in a recent paper~\cite{Lerose18}.

\bibliography{ising2}

\end{document}